\documentclass[12pt,preprint]{aastex}

\shorttitle{Scaling laws for jets}
\shortauthors{Huarte--Espinosa \& Mendoza}

\usepackage{booktabs}
\usepackage{amsmath,amssymb,amsbsy,amsfonts}
\begin{document}

\title{Simple scaling laws for astrophysical jets}
\author{M. Huarte--Espinosa$^1$ \& S. Mendoza$^2$}
\affil{$^1$ Churchill College, University of Cambridge, Cambridge CB3 ODS, 
       United Kingdom\\
       $^2$ Instituto de Astronom\'{\i}a, Universidad Nacional 
       Aut\'onoma de M\'exico, AP 70-264, Distrito Federal 04510,
       M\'exico}
\email{mh475@cam.ac.uk, sergio@astroscu.unam.mx}

\begin{abstract}
   The idea of a unified model for all astrophysical jets has been
considered for quite some time. We present here a few scaling
laws relevant to all type of astrophysical jets, analogous to those
of \citet{sams96} which are widely used for astrophysical black holes.
We use Buckingham's $ \Pi $ theorem of dimensional analysis to obtain a
family of dimensional relations among the physical quantities associated
to astrophysical jets.
\end{abstract}

% % Spanish abstract - leave blank and it will be translated by the
% % editors. 
% \resumen{ 
%   La idea de que exista un modelo unificado para todos los jets
% astrofisicos ha sido considerada extensamente desde hace tiemkpo.  En este
% art\'{\i}culo presentamos algunas relaciones de escala hidrodin\'amicas que
% son relevantes para todo tipo de jets astrof\'{\i}sicos y que son
% an\'alogas a aquellas de \citet{sams96}.  Utilizamos el teorema $ \Pi $ de
% Buckingham del an\'alisis dimensional para obtener una familia de
% relaciones dimensionales provenientes de las cantidades f\'{\i}sicas
% asociadas con jets.
% }

% % Keywords must be from the standard list and in alphabetical order. 
% \addkeyword{hydrodynamics}
% \addkeyword{galaxies: jets}
% \addkeyword{ISM: jets and outflows}
% \addkeyword{quasars: general}
% \addkeyword{gamma rays: bursts}


% Typeset article header
% \maketitle

\section{Introduction}

  Although the first report of an astrophysical jet was made by
\citet{curtis}, these objects were extensively studied much later with
radio astronomy techniques \citep{reber}.  Quasars, and radiogalaxies were
discovered and later gathered in a unified model which proposed a dusty
torus around the nucleus of the source \citep{antonucci85}. Years later,
some galactic sources showed similar features to the ones presented by
quasars and radiogalaxies, i.e. relativistic fluxes, a central engine,
symmetrical collimated jets, radiating lobes, and apparent superluminal
motions \citep[cf.][]{sunyaev91}. These objects are usually identified as
\( \mu \)--quasars. Optical and X-ray observations showed other
similar non--relativistic sources in the galaxy associated to H--H objects
\citep[cf.][]{pino04}. Lately, the strong explosions found in long Gamma Ray
Bursts (GRB) have been modelled as collapsars, in which a jet of a very short
lifetime is associated to the observed phenomena 
\citep[cf.][]{kulkarni99,castro99}.

% %%%%%%%%%%%%%%%%%% I M A G E %%%%%%%%%%%%%%%%%%%%%%%%%%%%%%%%%%%%%%
% \begin{figure}
%   \includegraphics[width=0.45\textwidth]{jets.eps}
% \caption{ 
%      Astrophysical jets are very common and exist in many different
%      sizes. On the left, extending $\sim 10^{5} \rm{pc}$, FR~I and FR~2
%      sources are shown. The upper right panel shows the micro--quasar
%      SS~433. It presents  relativistic fluxes and apparent superluminal
%      motions analogous to those in quasars. The lower right panel shows
%      jets associated to Herbig--Haro objects with lengths  $\sim 10^{-1}
%      - 10 \,\rm{pc}$. All  jets have a condensed (sometimes compact)
%      object accreting matter from their surroundings.  There is also
%      an accretion disc around the central condensed object and a pair
%      of symmetrical collimated jets that end up in radiating lobes.
%      The images were taken from \citet{bridle98}, \citet{paragi01} and
%      and hubblesite.org.
% }
% \label{fig1}
% \end{figure}
% %%%%%%%%%%%%%%%%%% I M A G E %%%%%%%%%%%%%%%%%%%%%%%%%%%%%%%%%%%%%%

  The similarities between all astrophysical jets, mainly those between
quasars and micro--quasars, and the scaling laws for black holes proposed
by \citet{sams96} and \citet{rees98} made us search for  the possible
existence of some scaling laws that may occur to astrophysical jets in
terms of very simple physical parameters, such as the magnetic field
associated to the accretion disc, accretion rate and mass of the central
object.  These sort of relations have been studied in a very different way
by \citet{heinz03,heinz05} giving scalings between the flux \( F_\nu \)
at a frequency \( \nu \) and the mass of the central object.

  The present work presents a few mathematical relations that appear
naturally as a consequence of dimensional analysis and Buckingham's \(
\Pi \) theorem.  We begin by considering some of the most natural physical
dimensional quantities that have to be included to describe some of the
physical phenomena related to all classes of jets.  We then calculate
the dimensional relations associated to these quantities.  Finally,
we briefly discuss these relations and their physical relevance to
astrophysical jets.

\section{Analysis}

  A complete set of quantities that must appear in the
physical description of a particular astrophysical jet is complicated.
However, there are some essential physical ingredients that must enter
into the description of the  problem.  To begin with, the mass \( M \)
of the central object must accrete material from its surroundings at an
accretion rate \( \dot{M} \).   Now, because gravity and magnetic fields
\( B \) are necessary in order to generate jets, Newton's constant
of gravity \( G \) and the velocity of light \( c \)  must also be
taken into account.  If in addition there are certain characteristic
quantities such as length \( l \), density \( \rho \) and velocity \( v \)
(e.g. these can be associated to the jet's length, the density of the
surrounding medium and the jet's ejection velocity respectively), then
the jet's kinetic luminosity or kinetic power \( L \) is an important
parameter that must be related in a general way to all these quantities
in the following manner:

\begin{equation}
  L = L(\dot{M},\, M,\, c,\, G,\,B,\, l,\, v,\, \rho).
 \label{ec4}
\end{equation}

\noindent Using Buckingham's $\Pi$ theorem of dimensional analysis
\citep{buckingham,sedov93} the following non--trivial dimensionless
parameters are found

\begin{alignat}{4}
  \Pi_1 &= \frac{ L }{  \dot{M} c^2 }, & \quad 
  \Pi_2 &= \frac{ G \dot{M} }{ c^3 }, & 
  \Pi_3 &= \frac{ B c^{1/2} M }{ \dot{M}^{ 3/2 } }, \label{pis} \\
  \Pi_4 &= \frac{ l \dot{M} }{ M c },  & \quad 
  \Pi_5 &= \frac{ \rho \ c^3 M^2 }{ \dot{M}^3 }. 
                                                        \notag
\end{alignat}

  The parameter $\Pi_2$ can be rewritten as

\begin{equation}   
 \Pi_2 = \left( \frac{G M}{ c^2} \right) \left( \frac{\dot{M}}{
 M} \right) \frac{1}{c}.
\notag 
\end{equation}

\noindent Since the quantity

\begin{equation}
 \tau := \frac{M}{\dot{M}},
 \label{tau}
\end{equation} 

\noindent has dimensions of time, according to Buckingham's \( \Pi \)
theorem it represents a characteristic time for our problem.
For the case we are analysing, it represents the time for which the central
object doubles its mass. Using equation~\eqref{tau} we can write \(
\Pi_2 \) as

\begin{equation}
 \Pi_2 = \frac{r_\text{s}}{ 2 \tau c},
\label{'lambda}
\end{equation}

\noindent where $r_\text{S} $ is the Schwarzschild radius.  This relation
naturally defines a length

\begin{equation}
 \lambda \sim c \tau,
 \label{lambda}
\end{equation} 

\noindent which can be thought of as the maximum possible length a
jet could have, since \( \tau \) is roughly an upper limit to the lifetime
of the source.

 On the other hand, from equation \eqref{pis} it is found that  

\begin{equation}
 \Pi_6 := \Pi_2^{3/2} \Pi_3 = \left( \frac{ G M
  }{ c^2 } \right)^{3/2}  \bigg/ \quad  \frac{  \sqrt{ M c^2 }  }{B}.   
 \label{ec5}
\end{equation}

\noindent The quantity \( G M / c^2 \) is of the order of the
gravitational radius, therefore the denominator of equation~\eqref{ec5}
defines a length \( r_\text{j} \) given by

\begin{equation}
  r_\text{j} \propto \frac{ M^{1/3} c^{2/3} }{ B^{2/3} } \approx
    10^2 \left( \frac{ M }{ \textrm{M}_\odot }
    \right)^{1/3}  \left( \frac{ B }{ 1 \rm{G} } \right)^{-2/3}
    \rm{pc}. 
\label{r_j}
\end{equation}

  From now on, we use typical values for different physical quantities
related to astrophysical jets as presented in Table~\ref{tabla}.
With these numerical values, it follows that extragalactic radio
sources and \( \mu \)--quasars are such that $r_\text{j} \propto 10^{4}
\, \rm{pc}$ and  $r_\text{j} \propto 10 \, \rm{pc}$ respectively.
These lengths are fairly similar to the associated length of their
corresponding jets.  Therefore, if we identify the length \( r_\text{j}
\) as the length of the jet, a constant of proportionality $ \sim 1 $
is needed in equation~\eqref{r_j}, and so

\begin{equation}
 r_\text{j} \approx 100 \left( \frac{ M }{ \textrm{M}_\odot } \right)^{1/3}  
 \left( \frac{ B }{ 1 \rm{G} } \right)^{-2/3} \rm{pc}. 
 \label{r_j=}
\end{equation}

%%%%%%%%%%%%%%%%%%%%%%%%%%%%  T A B L E  %%%%%%%%%%%%%%%%%%%%%%%%%%%
\begin{table*}[!t]
  % \setlength{\tabnotewidth}{0.9\textwidth}
    %The following stretches space between columns:
    % \setlength{\tabcolsep}{1.33\tabcolsep}
    % \tablecols{5}
%  \caption{Typical values for astrophysical jets\tabnotemark{a}.}
%  \label{tabla}
\begin{center}
   \begin{tabular*}{0.97\textwidth}{lcccccc}
     \toprule 
     Object  & AGN  & \( \mu \)--quasars  & H--H  & GRB  \\
     \midrule
     Size $\left[ \textrm{pc} \right]$  &$ \sim 10^5 $  &$ \lesssim
     10 $  &$ < \text{a few} $  &$ \sim 10^{-5} - 10^{-1} $  \\
     Luminosity $\left[ L_\odot \right]$  &$ 10^{7} - 10^{19} $  &
       $ < 10^{5}$  &$ 10^{1} - 10^{4}$  &$ 10^{21} $  \\
     Central mass $\left[ M_\odot \right]$  &$ 10^{6} - 10^{9} $  
     &$ 1 - 10 $  &$ < 10 $     & $ 1 - 10$  \\
     Lorentz factor $\left[ \Gamma \right] $  &$ 10 - 10^{3} $  &$ > 10
     $  &$ \lesssim 1.0000005 $    &$ 100 - 300  $  \\
     Magnetic field $\left[ \textrm{G} \right]$  &$ \sim 100 $  
     &$ \sim 100$  &$ < 200 $  &$\sim 10^{16}$ \\
     \bottomrule
%   \tabnotetext{a}{ Typical values for ...
%     \citet{blandford90,carilli96,lovelace96,shibata98,camenzind99,
%     meier02,wu02,smirnov03,uzdensky06}.}
  \end{tabular*} 
\end{center}
 \caption{Typical values for different physical quantities involved in the
     accretion--ejection phenomena of astrophysical jets.  
     All values of luminosities, central masses
     and velocities in terms of the Lorentz factor \( \Gamma \) 
     were taken from
     \citet{robson96,reipurth97,ferrari98,ford99,meier02,
     vilhu02,sunyaev03,calvet04,mendoza05,mirabel04,pino04}.  Magnetic
     field values were taken from
    \citet{blandford90,carilli96,lovelace96,shibata98,camenzind99,
    meier02,wu02,smirnov03,uzdensky06}.}
 \label{tabla}
\end{table*} 
%%%%%%%%%%%%%%%%%%%%%%%%%%%%  T A B L E  %%%%%%%%%%%%%%%%%%%%%%%%%%%

  Since equation~\eqref{ec5} is roughly the the Schwarzschild
radius $r_\text{S}$ divided by jet's length $r_\text{j}$ to the power 
\( 3/2 \), then 

\begin{equation} 
 \Pi_6 = \frac{ \left( B l^{3/2} \right) \left( G M^2/l
 \right)^{3/2} }{ (M c^2 )^2} << 1,
\label{r_j''}
\end{equation}
\noindent which in turn implies that
\begin{equation} 
 B << \frac{ c^4 }{ G^{3/2} M } \approx 10^{23} \left(
 M/\textrm{M}_\odot \right)^{-1} \, \rm{G}.
\label{r_j'2}
\end{equation}

 The right hand side of this inequality is the maximum upper limit
for the magnetic field associated to the accretion disc about the
central object. For ``extreme'' micro--quasars like SS~433 and GRB's
the magnetic field \( B \) reaches values $ \gtrsim 10^{16} \, \rm{G}$
\citep[cf.][]{meier02,uzdensky06}, so that this upper limit works better for
those objects.

   From equation~\eqref{pis} it follows that

\begin{equation}   
 \Pi_7 := \frac{\Pi_1}{\Pi_2 \Pi_3^2} = \frac{L \dot{M}}{
 B^2 M^2 G },
\notag 
\end{equation}   

\noindent and so

\begin{equation}   
 L \propto 10^{-7} \left( \frac{B}{1 \, \rm{G}} \right)^2  
 \left( \frac{M}{\textrm{M}_\odot} \right)^2  \left(
 \frac{\dot{M}}{\textrm{M}_\odot yr^{-1}}
 \right)^{-1} L_\odot.
\label{ls}
\end{equation}   

  For the case of quasars and \( \mu \)--quasars, using typical
values from Table~\ref{tabla}, it follows that 
the kinetic power $L \propto 10^{15} \, L_\odot$ and $L \propto  10^{8} \, 
L_\odot$ respectively.  In order to normalise it to the observed values, we
can set a constant of proportionality $\sim 10^{-6} $ in
equation~\eqref{ls}, and thus the jet power relation as a function of
the magnetic field, accretion rate and mass of the central object 
is given by

\begin{equation}   
  L \approx 10^{-13} \left( \frac{B}{ 1 \, \rm{G} } \right)^2  
  \left( \frac{M}{\textrm{M}_\odot} \right)^2  \left(
  \frac{\dot{M}}{\textrm{M}_\odot yr^{-1}}
  \right)^{-1} L_\odot.
  \label{ls'}
\end{equation}

\section{Discussion}

  According to the previous analysis astrophysical jets exist due to
a precise combination of electromagnetic, mechanical and gravitational
processes independently of the physical mechanisms behind the central engine.

  A trivial dimensionless parameter that is obtained using
Buckingham's \( \Pi \) theorem of dimensional analysis applied to
equation~\eqref{ec4} is the ratio \( v / c \).  With this, it is possible
to form another dimensionless parameter given by \( \Pi_2 \Pi_4^2 \Pi_5 /
\left( v / c \right)^2 \), which leads to the dimensionless quantity \(
\Lambda := \rho G l^2 / v^2 \) used by \citet{mendoza05} in order to
obtain a maximum length for an astrophysical jet.   As explained by
\citeauthor{mendoza05}, this maximum size is most probably determined by
the interaction of the jet and its cocoon with their surrounding environment,
leading to the generation of Kelvin--Helmholtz instabilities.

  Applying the results of equations \eqref{r_j} and \eqref{ls'} to GRB jets
with a canonical magnetic field \( B \sim 10^{16} \, \textrm{G} \),
leads to wrong output kinetic luminosities and typical sizes of GRB jets.
To correct these, proportionality factors of \( \sim 10^9 \) in the
jet's length \( r_j \) and of \( 10^{-5} \) in the kinetic power \( L \)
in equations \eqref{r_j} and \eqref{ls'} have to be used respectively.
The reason for this might be due to the fact that these jets have very
short life times and so, they hardly resemble a traditional steady jet.
Also, there might be some particular physical mechanisms that make
a dimensionless combination  changing the proportionality factors in
equations \eqref{r_j} and \eqref{ls'} in such a way that they give the
correct value needed for these class of jets.

  There are well known kinetic luminosities that appear in the literature
related  to the ejection of jets from different sources.  As an example,
in the \citet{blandfordpayne} model the luminosity takes the following form

\begin{equation}
  L = B_\text{p}^2 R^3 \Omega,
\label{bland-payne}
\end{equation}

\noindent where \( \Omega \) is the angular velocity of the poloidal
component  of the magnetic field \( B_\text{p} \) and \( R \)
is the size of the rotating region.  It is possible to obtain
equation~\eqref{bland-payne} with the model presented in this
article if we proceed as follows.  Let us include in the functional
relation~\eqref{ec4} an important parameter of the problem, namely an
angular velocity \( \Omega \), so that

\begin{equation}
  L = L(\dot{M},\, M,\, c,\, G,\,B,\, l,\, v,\, \rho,\, \Omega).
\label{lum-gen}
\end{equation}

\noindent It is then possible to build another dimensionless parameter \(
\Pi_7 \) given by

\begin{equation}
  \Pi_7 := \frac{ \Omega M  }{ \dot{M} }.
\label{pi7}
\end{equation}

  The \citet{blandfordpayne} kinetic luminosity is obtained using
equations \eqref{pis} and \eqref{pi7} with the introduction of a new
dimensionless parameter \( \Pi_8 := \Pi_3^2 \, \Pi_4^3 \, 
\Pi_7 \, / \, \Pi_1 \). 

  In the same manner, let us define a new dimensionless parameter 
\( \Pi_9 := \Pi_3^2 \, \Pi_4^4 \, \Pi_7^2 \, / \, \Pi_1 \) which 
reproduces the general dimensional shape
of the kinetic luminosity of the \citet{blandfordznajek} model, 
giving \citep{meier02}

\begin{equation}
  L \propto \frac{ 1 }{ c } B^2 R^4 \Omega^2.
\label{blandford-znajek}
\end{equation}

\noindent The constant of proportionality in this equation has a value
of \( \sim \, 0.1 \, \)--\( \, 0.03 \) depending on the geometry of the
problem.
  
  Of all our results, it is striking the fact that the jet power is
inversely proportional to the accretion rate associated with it.  This
is probably due to the following.  For a fixed value of the
mass of the central object (in any case, for the time that accretion
takes place, the mass of the central object does not increase too much)
when the accretion mass rate increases, the magnetic field lines
anchored to the plasma tend to pack up and thus, the field's intensity
increases in such a way as to get the correct result given by
equation~\eqref{ls'}.

  All the results presented in this article are in agreement with powerful
extragalactic jets and jets associated with \( \mu \)--quasars and
Herbig--Haro objects.  They are also in agreement with jets associated to
long gamma--ray bursts if proportionality factors are adjusted for the
corresponding physical quantities.

  Our main result, that the kinetic luminosity of the jet as a function of the
magnetic field, the accretion rate and the mass of the central object is
of general validity, independently of whether the same physical mechanism
produces jets from galactic sources (i.e. Herbig--Haro, \( \mu \)-quasar
and possibly jets associated to Gamma Ray Bursts) or extragalactic ones.
Indeed, if the physics behind the accretion--ejection mechanism that
occurs in jets works differently for each class of jets,
then the luminosity output will be given by  different mathematical
expressions.  For example, the \citet{blandfordpayne} and
\citet{blandfordznajek} models, are given by equations
\eqref{bland-payne} and \eqref{blandford-znajek} respectively. However,
for these two particular cases, we have proven above that if the Luminosity
is a general function given by equation \eqref{lum-gen}, then both cases lead to
the luminosity relation \eqref{ls}.  On the other hand, if
the physical mechanism that generates jets at all scales is the same,
relation \eqref{ls} is also of general validity if the
luminosity is a function described by \eqref{ec4}, or equivalently by
\eqref{lum-gen}.

\section{Acknowledgements}

  We would like to thank S. Setiawan for useful discussions about jet
power and their association with gravitational effects.  S.M. gratefully
acknowledges support from DGAPA (IN119203) at Universidad Nacional
Aut\'onoma de M\'exico (UNAM).  M.H.E acknowledges support from CONACyT
(196898/217314).

% %%%%%%%%%%%%%%%%
% % BIBLIOGRAPHY %
% %%%%%%%%%%%%%%%%
\bibliography{jetty}
\bibliographystyle{apj}
% %%%%%%%%%%%%%%%%

\end{document}